\documentclass[]{aastex631}

\shorttitle{}
\shortauthors{Warnhofer et al.}
\graphicspath{{./}{figures/}}

\begin{document}

\title{Multi-View Deep Learning for Imaging Atmospheric Cherenkov Telescopes}

\author[0009-0004-4883-4079]{Hannes Warnhofer}
\affiliation{Erlangen Centre for Astroparticle Physics (ECAP), Friedrich-Alexander-Universit{\"a}t Erlangen-N{\"u}rnberg, Nikolaus-Fiebiger-Str. 2, D 91058 Erlangen, Germany\\}

\author[0000-0001-5516-1205]{Samuel T. Spencer}
\affiliation{Erlangen Centre for Astroparticle Physics (ECAP), Friedrich-Alexander-Universit{\"a}t Erlangen-N{\"u}rnberg, Nikolaus-Fiebiger-Str. 2, D 91058 Erlangen, Germany\\}
\affiliation{Department of Physics, Clarendon Laboratory, Parks Road, Oxford, OX1 3PU, United Kingdom}

\author[0000-0003-3631-5648]{Alison M.W. Mitchell}
\affiliation{Erlangen Centre for Astroparticle Physics (ECAP), Friedrich-Alexander-Universit{\"a}t Erlangen-N{\"u}rnberg, Nikolaus-Fiebiger-Str. 2, D 91058 Erlangen, Germany\\}



\begin{abstract}
\noindent
This research note concerns the application of deep-learning-based multi-view-imaging techniques to data from the H.E.S.S. Imaging Atmospheric Cherenkov Telescope array. We find that the earlier the fusion of layer information from different views takes place in the neural network, the better our model performs with this data. Our analysis shows that the point in the network where the information from the different views is combined is far more important for the model performance than the method used to combine the information. 

\end{abstract}

\keywords{Gamma-ray telescopes (634) --- Astronomy image processing (2306)}


\section{Introduction}
\noindent
Imaging Atmospheric Cherenkov Telescopes (IACTs) observe the very-high-energy gamma-ray sky by measuring the properties of air showers produced when such photons interact with Earth's atmosphere. Rejection of a large hadronic background is key to the sensitivity of IACTs, with deep-learning-based methods of event classification being a promising approach for this task \citep{SHILON201944,sam,Glombitza_2023}. Stereoscopic image analysis is critical for IACT performance, but combining this with deep learning techniques is an unsolved problem. Early works simply summed the images from multiple cameras together (e.g. \citet{mangano}), but this is known to dilute features present in the individual images, and was shown by \citet{SHILON201944} to not work robustly on IACT data for events with a telescope multiplicity $>3$. \citet{SHILON201944} suggested an approach based on the hybridisation of Convolutional Neural Network (CNN) and Recurrent Neural Network (RNN) layers. However, their underlying argument that such network architectures allowed the networks to exploit the air shower to telescope distance was questioned in \citet{ari} and \cite{sam}. In this note, we explore recent multi-view-imaging methods from computer science \citep{seeland2021} for this task.

\section{Methods}
\noindent
We aim to build multi-view CNN models for binary IACT event classification that utilise different methods of fusing layer information. Following \citet{seeland2021}, the base for these multi-view CNNs is a single-view CNN which takes a single image as input. The input is fed through 7 blocks, each consisting of a dropout layer, a 2D convolutional layer and a 2D maximum pooling layer, with only the number of filters in the convolution changing from block-to-block. The output of the last block is flattened and fed through a fully connected (FC) layer with 1024 filters. Then another 50\,\% dropout is applied before entering the classification layer with a sigmoid activation function. All the multi-view CNNs consist of four parallel CNNs with the same architecture as the single-view CNN, until the fusion operation is applied, at the so-called ``fusion location''. The output of the fusion layer is then fed into the remaining part of the single-view CNN.\\
\par
\noindent
A model architecture based on ResNet-50 resulted in strong overfitting of the models, therefore we use a custom CNN for the single view model, as \citet{seeland2021} claim their fusion methods should be agnostic to the architecture used. Similarly, freezing the single-view CNN weights following \citet{seeland2021} resulted in poor performance, and is hence not used here. For our analysis, we use the same dataset and cuts described in \cite{Glombitza_2023} for simulated data of the CT1-4 telescopes of the High Energy Stereoscopic System (H.E.S.S.) IACT array. We consider the three fusion strategies of Early, Late and Score Fusion \citep{seeland2021}.\\
\par
\noindent
For \textbf{Early Fusion}, the outputs of the second block in the base CNN (the second max pooling layers) are fused and the output of the fusion layer is fed into the remaining three blocks. These are then flattened and fed through a FC layer, a dropout layer and the classification layer. Three different methods of fusing the information at this stage are analysed. Two fusion methods stack the layer outputs along a new axis, and for each spatial position of the final feature map, the information of the four stacked feature maps is reduced. \textit{Early Max Fusion} applies a max pooling operation along the stacked feature maps, leaving the maximum value of the four layers for each spatial position. \textit{Early Conv Fusion} uses a 1x1 convolutional operation instead. Lastly, the \textit{Early Concat Fusion} method concatenates the four layer outputs along the channel axis. This results in a new layer with four times the amount of filters, which is subsequently fed through the remaining CNN.\\
\par
\noindent
\textbf{Late Fusion} takes place after the last FC layer before the classifier. For \textit{Late FC Fusion}, the output feature vectors of the dense layers are concatenated along the channel axis before being fed through a FC layer. For \textit{Late Max Fusion} the feature vectors are fused analogously to \textit{Early Max Fusion}, by stacking the four layer outputs before applying a max pooling operation along the new axis.\\
\par
\noindent
For \textbf{Score Fusion} the images of each view are fed through the complete single-view CNN and the final scores of the classification layer are combined. Three approaches of score fusion are analysed: \textit{Score Mean Fusion} and \textit{Score Prod Fusion} store the mean and the product of the scores respectively, whereas \textit{Score Max Fusion} applies a maximum pooling operation.\\
\par
\noindent
The \textit{Adam} optimizer \citep{adam} was used for training, and an early stopping callback monitoring the validation loss was applied. 100,000 events were used, 80\% for training and 20\% for validation. Network architecture diagrams and the code are available at \url{https://github.com/hanneswarnhofer/multiview-cnn-fusion-iact.git} (\url{https://doi.org/10.5281/zenodo.10868020}).\\

\section{Results and Conclusions}
\noindent
The models were evaluated using the receiver operating characteristic (ROC) curve, and the accuracy over the respective overall event size (as a proxy for performance variation with event energy). The event size is given by the sum of all pixel values present after selection cuts.\\
\par
\begin{figure*}
\includegraphics[width=\columnwidth]{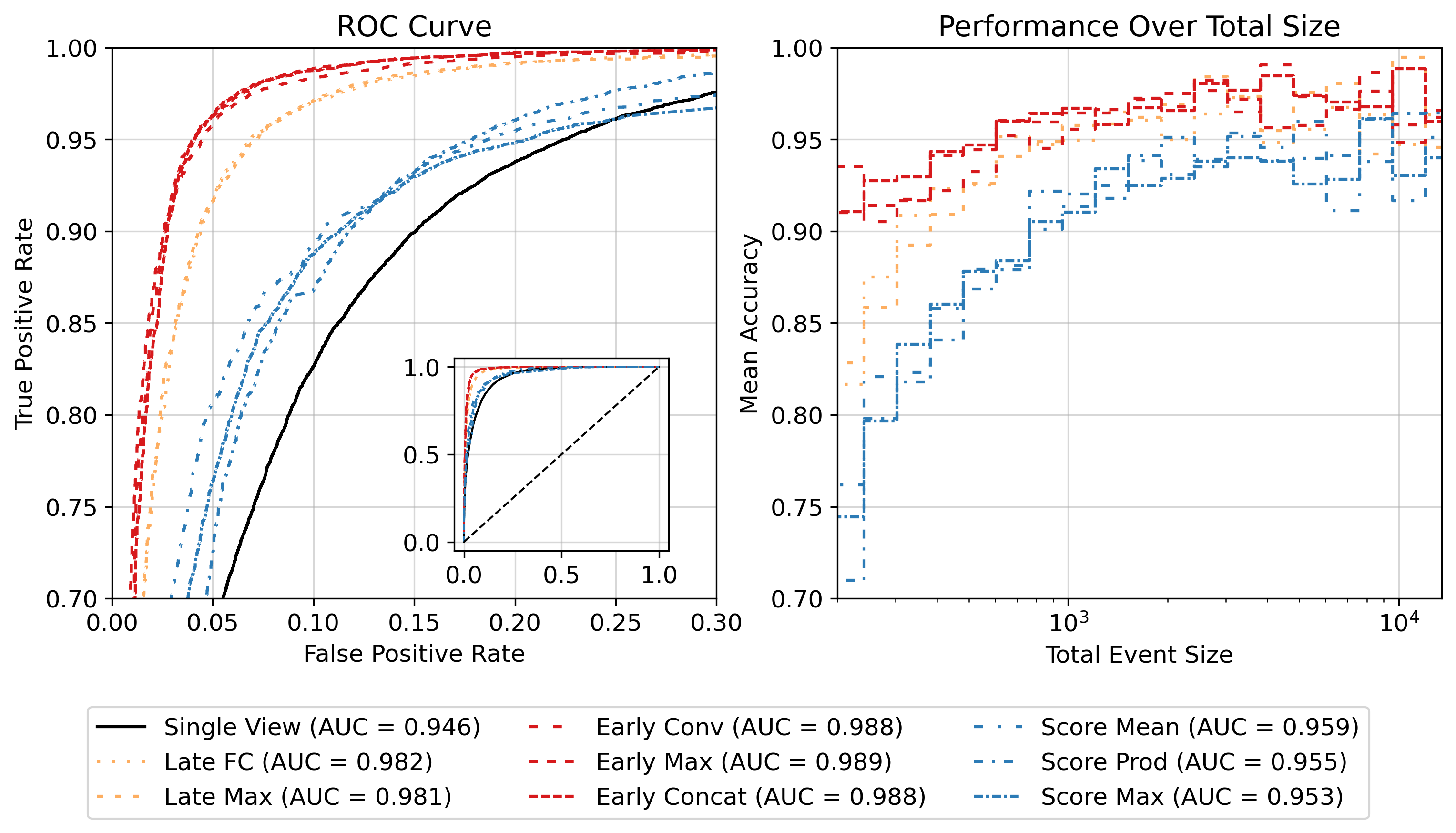}
\caption{ROC curves and performance over size of the differently fused models.}
\label{fig:main}
\end{figure*}
\noindent
Figure \ref{fig:main} shows that the multi-view models achieve superior performance compared to the single-view case, although the Score Fusion models only marginally improve on the single-view CNN. We find the fusion layer location influences model performance more than the fusion method. Particularly for Early and Late Fusion, there is only a minimal effect of the method on the area under curve (AUC) score, varying up to 0.001 for a fixed fusion location. For Late Fusion, the AUC score is $0.982$ and for Early Fusion $0.989$. For Score Fusion, the variation in the ROC curves of the different methods is higher, whilst in general with earlier fusion the model performance improves. Early Fusion after the fourth block was also investigated, with an average AUC of $0.987$, lower by 0.002 compared to Early Fusion after the second block. We find that all models perform worse for lower sizes, as expected  \citep{Glombitza_2023}. For higher sizes, the different fusion methods show similar accuracy, with score fusion even outperforming early fusion in some size bins.\\
\par
\noindent
Overall, we find that earlier view fusion strategies work best for IACT data,  contrary to \citet{seeland2021} who concluded that, although dataset dependent, later fusion typically led to increased accuracy. The fact that the spatial scales and image gradients present in IACT images differ to conventional photographic datasets could explain this discrepancy. IACT images being innately greyscale may also be a factor, as most computer science datasets use 3 channels. Our results ultimately support the statement that there is no universally superior method for multi-view imaging across all datasets, and this analysis should be repeated when applying these view fusion strategies to new stereoscopic data.\\


\section{Acknowledgements}
\noindent
We thank \citet{Glombitza_2023} for sharing their data, and thank the H.E.S.S. collaboration for permission to use their simulation models. This work is supported by the Deutsche Forschungsgemeinschaft (DFG, German Research Foundation) – Project Number 452934793. 

\bibliography{hessmlanalysis}{}
\bibliographystyle{aasjournal}



\end{document}